\documentclass[12pt]{article}
\usepackage{amsfonts}

\newcommand{\eqdef}{\stackrel{\rm def}{=}}

\newcommand{\ignore}[1]{}

\begin{document}

\baselineskip=20pt
\renewcommand{\theequation}{\arabic{section}.\arabic{equation}}

\newfont{\elevenmib}{cmmib10 scaled\magstep1}
\newcommand{\preprint}{
    \vspace*{-20mm}\begin{flushleft} Version: 29 August 2007 \end{flushleft}
    \begin{flushleft}
      \elevenmib Yukawa\, Institute\, Kyoto\\
    \end{flushleft}\vspace{-1.3cm}
    \begin{flushright}\normalsize  \sf
      YITP-07-42\\
      July 2007
    \end{flushright}}
\newcommand{\Title}[1]{{\baselineskip=26pt
    \begin{center} \Large \bf #1 \\ \ \\ \end{center}}}
\newcommand{\Author}{\begin{center}
    \large \bf  Ryu Sasaki \end{center}}
\newcommand{\Address}{\begin{center}
     Yukawa Institute for Theoretical Physics,\\
      Kyoto University, Kyoto 606-8502, Japan
    \end{center}}
\newcommand{\Accepted}[1]{\begin{center}
    {\large \sf #1}\\ \vspace{1mm}{\small \sf Accepted for Publication}
    \end{center}}

\preprint
\thispagestyle{empty}
\bigskip\bigskip\bigskip

\Title{Quasi Exactly Solvable Difference Equations}
\Author

\Address
\vspace{1cm}

\begin{abstract}
Several explicit examples of {\em quasi exactly solvable\/} `discrete' quantum
mechanical Hamiltonians are derived by deforming the well-known exactly solvable
Hamiltonians of one degree of freedom. 
These are difference analogues of the well-known quasi exactly solvable systems,
the harmonic oscillator (with/without the centrifugal potential) deformed by a
sextic potential and the $1/\sin^2x$ potential deformed by a $\cos2x$ potential.
They
have a finite number of exactly calculable eigenvalues and
eigenfunctions.
\end{abstract}

\section{Introduction}
\label{intro}
Exactly solvable and Quasi Exactly Solvable (QES) quantum mechanical
systems have played a very important role in modern physics.
The former, the exactly solvable systems, are quite well-known.
In the Schr\"odinger picture, if all the eigenvalues and corresponding
eigenfunctions are known, the system is exactly solvable.
Plenty of such systems are known, for example,
the P\"oschl-Teller and the Morse potential on top of the best-known
harmonic oscillator and the coulomb potential \cite{susyqm} 
for degree one cases and the Calogero-Sutherland systems \cite{cal,sut,kps}
for many degrees of freedom cases.
Recently the exact Heisenberg operator solutions and the corresponding
annihilation-creation operators are constructed for most of the degree one
exactly solvable quantum mechanics \cite{os7} and for the multi-particle
Calogero systems \cite{os9}.
The notion of exact solvability was generalised to the so-called `discrete'
quantum mechanics, in which the Schr\"odinger equation is a {\em difference\/}
equation in stead of differential.
The difference analogues of the  Calogero-Sutherland systems were constructed by
Ruijsenaars-Schneider-van Diejen \cite{RS,vD}. 
The difference equation analogues of the equations determining the Hermite,
Laguerre and Jacobi polynomials were derived within the Askey-scheme of
hypergeometric orthogonal polynomials \cite{And-Ask-Roy,koeswart}.
Later they were reformulated as Hamiltonian dynamics with 
{\em shape-invariance\/}
\cite{genden} by Odake-Sasaki \cite{os4os5}.

In contrast, the latter, Quasi Exactly Solvable (QES) systems have a short history
and less known. If a finite number of exact eigenvalues and eigenfunctions are
known, the system is QES \cite{Ush}. 
Since the number  of exactly solvable states can be chosen as
large as wanted, a QES system could be used as a good alternative to
an exactly solvable system. Several one
degree of freedom QES systems are listed in
\cite{Ush,turb} and multi-particle QES systems were first constructed by
Sasaki-Takasaki
\cite{st1} as deformation of Inozemtsev- Calogero-Sutherland systems,  which was
followed by
\cite{turb2}. 

In the present paper we derive several explicit examples of QES difference
equations as deformation of {\em exactly solvable\/} `discrete'
quantum mechanics \cite{os4os5}. They are difference analogues of the well-known
quasi exactly solvable systems, the harmonic oscillator (with/without the
centrifugal potential) deformed by a sextic potential and the $1/\sin^2x$
potential deformed by a
$\cos2x$ potential. 

This paper is organised as follows. In the next section, the deformation method to
obtain a QES from an exactly solvable system is explained in some
detail by taking two well-known examples of the ordinary quantum mechanical
QES systems.
Then two corresponding difference equation analogues are derived.
Section \ref{Examples} provides three more explicit examples. 
The final section is for a summary and comments.

\section{Quasi Exactly Solvable Deformation}
\label{QESdef}
\setcounter{equation}{0}

There are many different ways of deriving  QES Hamiltonians for 
ordinary quantum mechanics \cite{Ush,turb}. However, to the best of our knowledge, 
a very limited number of explicit examples of QES {\em difference\/} equations are
known  in connection with $U_q(sl(2))$ \cite{Wiegmann}. 
In these examples, quantum wavefunctions are related to those defined on {\em
discrete lattice points\/} only.  In the present paper we
present several explicit  examples of QES `discrete' quantum mechanical
Hamiltonians of one degree of freedom, whose wavefunctions are continuous
functions of $x$ as in the ordinary quantum mechanics. They are
obtained by deforming {\em exactly solvable\/} `discrete' quantum mechanical
Hamiltonians
\cite{os4os5}, which have the Askey-scheme of hypergeometric orthogonal polynomials
\cite{And-Ask-Roy,koeswart} as part of the eigenfunctions; the Meixner-Pollaczek,
continuous Hahn, continuous dual Hahn, Wilson and Askey-Wilson polynomials. This
deformation method was first applied by Sasaki and Takasaki \cite{st1}  to derive
multi-particle QES based on the Inozemtsev models.

For illustrative purposes, we will explain the deformation method 
for the two well-known examples of degree one QES systems 
in ordinary quantum mechanics in the next subsection. These examples are the sextic ($x^6$) potential added to the harmonic oscillator ($x^2$) potential, and another a $\cos2x$ potential added to the exactly solvable $1/\sin^2x$ potential. The same method is applied in  subsection \ref{dqmex} to derive the first two examples of QES Hamiltonian in `discrete' quantum mechanics 
corresponding to the the sextic potential deformation. 
The rest of the examples are given in 
section \ref{Examples}.

\subsection{Ordinary Quantum Mechanics}
\label{oqmex}
\subsubsection{Harmonic Oscillator Deformed by Sextic Potential}

A best-known example of QES Hamiltonian, 
the harmonic oscillator plus a sextic ($x^6$)
potential is given succinctly by
\begin{equation}
\mathcal{H}\eqdef-\frac{d^2}{dx^2}+\left(\frac{dW}{dx}\right)^2
+\frac{d^2W}{dx^2}+\alpha_\mathcal{M}(x),\quad
\alpha_\mathcal{M}(x)\eqdef-2a\mathcal{M} x^2, \quad
\mathcal{M}\in
\mathbb{N}.
\label{sextic}
\end{equation}
Here we tentatively call the last term of the above Hamiltonian,
$\alpha_\mathcal{M}(x)$,  the compensation term. Throughout this paper
we adopt the unit system
$2m=\hbar=1$. The real prepotential $W$ is a deformation of that for the
harmonic  oscillator $W_0\eqdef -bx^2/2$, with $a$ being the deformation parameter:
\begin{equation}
W=W(x)\eqdef-\frac{a}{4}x^4+W_0=-\frac{a}{4}x^4-\frac{b}{2}x^2,\qquad a,\ b>0. 
\end{equation}
By the similarity transformation in terms of the {\em pseudo ground state\/} 
wavefunction
$\phi_0(x)\eqdef e^{W(x)}$, we obtain
\begin{eqnarray}
  \tilde{\mathcal{H}}\eqdef
\phi_0^{-1}\circ\mathcal{H}\circ\phi_0&=&
  -\frac{d^2}{dx^2}
  -2\frac{dW(x)}{dx}\frac{d}{dx}-2a\mathcal{M}x^2,
\label{ordsim}\\
&=&-\frac{d^2}{dx^2}+\left(2ax^3+2bx
\right)\frac{d}{dx}-2a\mathcal{M}x^2.
\end{eqnarray}
In the absence of the compensation term $\alpha_\mathcal{M}(x)$, $\phi_0(x)$
is actually a ground state wavefunction, therefore it has no node and it is square
integrable.
Another characterisation of the {\em pseudo ground state\/} 
wavefunction $\phi_0$ is that it is annihilated by the operator $A$ 
which factorises the main part of the Hamiltonian (\ref{sextic}):
\begin{eqnarray}
A\phi_0(x)&=0,& \quad A\eqdef -\frac{d}{dx}+\frac{dW(x)}{dx},\quad 
A^\dagger\eqdef \frac{d}{dx}+\frac{dW(x)}{dx},\\
\mathcal{H}&=&A^\dagger A-2a\mathcal{M}x^2.
\end{eqnarray}
The action of the Hamiltonian $\tilde{\mathcal{H}}$ (\ref{ordsim}) 
on monomials of $x$ reads
\begin{equation}
\tilde{\mathcal{H}}\,x^n=\left\{\begin{array}{ll}
-n(n-1)x^{n-2}+2nbx^n+2a(n-\mathcal{M})x^{n+2},& n\leq \mathcal{M}-2,\\[4pt]
-\mathcal{M}(\mathcal{M}-1)x^{\mathcal{M}-2}+2\mathcal{M}b x^\mathcal{M},& 
n= \mathcal{M}.
\end{array}\right.
\end{equation}
Since the parity is conserved,
it is now obvious that $\tilde{\mathcal{H}}$ keeps the polynomial space
$ {\mathcal V}_{\mathcal M}$ invariant, 
\begin{eqnarray}
\tilde{\mathcal{H}}\,{\mathcal V}_{\mathcal M}&\subseteq&
{\mathcal V}_{\mathcal M},
\label{subspac1}\\[4pt]
    {\mathcal V}_{\mathcal M}
&\eqdef&
\left\{\begin{array}{ll}
\mbox{Span}\left[1,x^2,\ldots,x^{2k},\ldots,x^{\mathcal 
    M}\right],& \mathcal{M}:\mbox{even},\\[4pt]
\mbox{Span}\left[x,x^3,\ldots,x^{2k+1},\ldots,x^{\mathcal 
    M}\right],& \mathcal{M}:\mbox{odd},
\end{array}\right.
    \label{eq:V0def}
\end{eqnarray}
and that $\tilde{\mathcal{H}}$ is a {\em tri-diagonal\/}  matrix.
Thus we can obtain a finite number of exact eigenvalues and eigenfunctions
of the sextic potential Hamiltonian (\ref{sextic}) in the form:
\begin{equation}
{\mathcal H}\phi={\mathcal E}\phi,\quad
\phi(x)=\phi_0(x)P_{\mathcal M}(x),\quad 
P_{\mathcal M}\in {\mathcal V}_{\mathcal M},\Longleftrightarrow
\tilde{\mathcal H}P_{\mathcal M}={\mathcal E}P_{\mathcal M},
\label{eigfun}
\end{equation}
by the {\em diagonalisation of a finite dimensional Hamiltonian matrix\/} 
$\tilde{\mathcal H}$ (\ref{ordsim}) with
\begin{equation}
\mbox{dim}{\mathcal V}_{\mathcal M}=\left\{
\begin{array}{ll}
{\mathcal M}/2+1,& {\mathcal M}: \mbox{even},\\
({\mathcal M}+1)/2,& {\mathcal M}: \mbox{odd}.
\end{array}
\right.
\label{Vdimform}
\end{equation}
Since $\mathcal{H}$ is obviously hermitian (or self-adjoint), 
all the eigenvalues are real and eigenfunctions belonging to different
eigenvalues are orthogonal with each other. 
In other words, two polynomial solutions $P_{\mathcal M}$ and $P'_{\mathcal M}$
are orthogonal with respect to the weight function $\phi^2_0(x)$.
The square integrability of all
the eigenfunctions of the above form
(\ref{eigfun}) $\int_{-\infty}^\infty\phi^2(x)dx<\infty$ is obvious. The {\em
true\/} ground state wave function has the form  (\ref{eigfun}) with the lowest
eigenvalue, say $\mathcal{E}_0$ and it has no node due to the oscillation theorem.

\subsubsection{$1/\sin^2x$ Potential Deformed by $\cos2x$ Potential}
\label{cos2mod}
Another well-known example of quasi exactly solvable system is the exactly
solvable $1/\sin^2x$  potential ($W_0\eqdef g\log\sin x$) deformed by a 
$\cos2x$ potential. The Hamiltonian has the same form as (\ref{sextic})
with only the prepotential $W(x)$ and the compensation term 
$\alpha_\mathcal{M}(x)$ different:
\begin{eqnarray}
\mathcal{H}&\eqdef&-\frac{d^2}{dx^2}+\left(\frac{dW}{dx}\right)^2
+\frac{d^2W}{dx^2}+\alpha_\mathcal{M}(x),\quad
\alpha_\mathcal{M}(x)\eqdef 4a\mathcal{M}
\sin^2x, \quad
\mathcal{M}\in
\mathbb{N},
\label{cos2}\\
W&\eqdef&\frac{a}{2}\cos2x+W_0=\frac{a}{2}\cos2x+g\log \sin x,
\quad g>0,\quad 0< x<
\pi.
\end{eqnarray}
Again by the similarity transformation in terms of the {\em pseudo ground state\/} 
wavefunction
$A\phi_0=0$, $\phi_0(x)\eqdef e^{W(x)}$,  we obtain
\begin{eqnarray}
  \tilde{\mathcal{H}}\eqdef
\phi_0^{-1}\circ\mathcal{H}\circ\phi_0&=&
  -\frac{d^2}{dx^2}
  -2\frac{dW(x)}{dx}\frac{d}{dx}+4a\mathcal{M}\sin^2x,
\label{ordsim2}\\
&=&-\frac{d^2}{dx^2}+\left(2a\sin2x-2g\cot x
\right)\frac{d}{dx}+4a\mathcal{M}\sin^2x.
\end{eqnarray}
Needless to say $\phi_0$ has no node or singularity and it is square
integrable $\int_0^\pi \phi_0^2(x)dx<\infty$.
The action of the Hamiltonian $\tilde{\mathcal{H}}$ (\ref{ordsim2}) 
on monomials of $\sin x$ reads
\begin{eqnarray}
&&\quad\tilde{\mathcal{H}}\,\sin^n\!x\\
&&\hspace*{-8pt}=\left\{\begin{array}{ll}
\!\!-n(n-1+2g)\sin^{n-2}\!x+n(n+2g+4a)\sin^n\!x+4a(\mathcal{M}-n)\sin^{n+2}\!x,&
n\leq
\mathcal{M}-2,\\[4pt]
\!\!
-\mathcal{M}(\mathcal{M}-1+2g)\sin^{\mathcal{M}-2}\!x+\mathcal{M}(\mathcal{M}+2g
+4a)
\sin^\mathcal{M}\!x,&  n= \mathcal{M}.
\end{array}\right.\nonumber
\end{eqnarray}
Since the parity is conserved,
it is now obvious that $\tilde{\mathcal{H}}$ keeps the polynomial space
$ {\mathcal V}_{\mathcal M}$ invariant, 
\begin{eqnarray}
\tilde{\mathcal{H}}\,{\mathcal V}_{\mathcal M}&\subseteq&
{\mathcal V}_{\mathcal M},
\label{subspac2}\\[4pt]
    {\mathcal V}_{\mathcal M}
&\eqdef&
\left\{\begin{array}{ll}
\mbox{Span}\left[1,\sin^2\!x,\ldots,\sin^{2k}\!x,\ldots,
\sin^{\mathcal M}\!x\right],& \mathcal{M}:\mbox{even},\\[4pt]
\mbox{Span}\left[\sin x,\sin^3\!x,\ldots,\sin^{2k+1}\!x,\ldots,
\sin^{\mathcal  M}\!x\right],& \mathcal{M}:\mbox{odd},
\end{array}\right.
    \label{eq:V0def2}\\
&&\mbox{dim}{\mathcal V}_{\mathcal M}=\left\{
\begin{array}{ll}
{\mathcal M}/2+1,& {\mathcal M}: \mbox{even},\\
({\mathcal M}+1)/2,& {\mathcal M}: \mbox{odd},
\end{array}
\right.
\label{Vdimform2}
\end{eqnarray}
and that $\tilde{\mathcal{H}}$ is again a {\em tri-diagonal\/}  matrix.
Thus we can obtain a finite number  (dim$\mathcal{V}_{\mathcal M}$) of exact
eigenvalues and eigenfunctions in the same way as in  the sextic potential
Hamiltonian (\ref{sextic}) case.

\subsection{`Discrete' Quantum Mechanics}
\label{dqmex}
\subsubsection{Difference Equation Analogue of Harmonic Oscillator Deformed by Sextic
Potential I}
\label{difex1}

A difference analogue of the sextic potential Hamiltonian (\ref{sextic}) is
\begin{eqnarray}
  \mathcal{H}&\eqdef&\!\!\sqrt{V(x)}\,e^{-i\partial_x}\sqrt{V(x)^*}
  +\!\sqrt{V(x)^*}\,e^{i\partial_x}\sqrt{V(x)}
  -(V(x)+\!V(x)^*)\!+\!\alpha_\mathcal{M}(x),
  \label{H1}\\[4pt]
&=&A^\dagger A+\alpha_\mathcal{M}(x),\hspace{30mm} \alpha_\mathcal{M}(x)\eqdef
2\mathcal{M} x^2,\label{factform}\\ A^{\dagger}
  &\eqdef&\sqrt{V(x)}\,e^{-\frac{i}{2}\partial_x}
  -\sqrt{V(x)^*}\,e^{\frac{i}{2}\partial_x},\quad
 A \eqdef e^{-\frac{i}{2}\partial_x}\sqrt{V(x)^*}
  -e^{\frac{i}{2}\partial_x}\sqrt{V(x)},\label{defAAdag}\\
 V(x)&\eqdef&(a+i x)(b+i x)V_0(x),\quad 
V_0(x)\eqdef c+ ix,\quad a,b,c\in\mathbb{R}_+.
\end{eqnarray}
Here as usual $V(x)^*$ is the complex conjugate of $V(x)$.
If $V$ is replaced by $V_0$ and the last term in (\ref{H1}),
$\alpha_\mathcal{M}(x)$, is
removed, $\mathcal{H}$ becomes the exactly solvable Hamiltonian of a  difference
analogue of the harmonic oscillator, or the {\em deformed harmonic
oscillator\/} in `discrete' quantum mechanics \cite{os4os5}.
Its eigenfunctions consist of the Meixner-Pollaczek polynomial, 
which is a deformation of the Hermite polynomial \cite{os4os5,degruij}. 
The quadratic polynomial factor $(a+i x)(b+i x)$ can be
considered as multiplicative deformation, 
although the parameters $a$, $b$ and $c$ are
on the equal footing. On the other hand one can consider it as a multiplicative
deformation by a linear polynomial in $x$:
\[
V(x)=(a+i x)V_{01}(x),\quad V_{01}(x)\eqdef(b+i x)(c+i x),
\]
with $V_{01}$ describing another difference version of an exactly 
solvable  analogue 
of the harmonic oscillator \cite{os4os5}. Its eigenfunctions consist of the 
continuous Hahn polynomial.

Next let us introduce the similarity transformation in terms of 
the {\em pseudo ground state\/} wavefunction $\phi_0(x)$:
\begin{eqnarray}
\phi_0(x)&\eqdef&\sqrt{\Gamma(a+i x)\Gamma(a-i x)
\Gamma(b+i x)\Gamma(b-i x)\Gamma(c+i
x)\Gamma(c-i x)},\\
 \tilde{\mathcal{H}}&\eqdef&
  \phi_0^{-1}\circ\mathcal{H}\circ\phi_0
  =V(x)\left(e^{-i\partial_x}-1\right)+V(x)^*\left(e^{i\partial_x}-1\right)
  +2\mathcal{M} x^2.
  \label{tilH}
\end{eqnarray}
It is obvious that $\phi_0$ has no node and that it is square integrable.
As in the ordinary quantum mechanics, $\phi_0(x)$ is annihilated by the  $A$
operator (\ref{defAAdag})
\begin{equation}
0=A\,\phi_0(x)=\left(e^{-\frac{i}{2}\partial_x}\sqrt{V(x)^*}
  -e^{\frac{i}{2}\partial_x}\sqrt{V(x)}\right)\phi_0(x).
\end{equation}
It is rather trivial to verify the action of the Hamiltonian $\tilde{\mathcal{H}}$
(\ref{tilH})  on monomials of $x$:
\begin{equation}
\tilde{\mathcal{H}}\,x^n=\left\{\begin{array}{lll}
\sum_{j=0}^{[n/2+1]}a_{n,\,j}x^{n+2-2j},& n\leq \mathcal{M}-2,&
a_{n,\,j}\in\mathbb{R},\\[6pt]
\sum_{j=0}^{[\mathcal{M}/2]}a'_{n,\,j}x^{\mathcal{M}-2j},&  n=
\mathcal{M},&a'_{n,\,j}\in\mathbb{R}.
\end{array}\right.
\label{Hactx}
\end{equation}
Here $[m]$ is the standard Gauss' symbol denoting the greatest integer  not
exceeding or equal to
$m$. Since the parity is conserved,
$\tilde{\mathcal{H}}$ keeps the polynomial space
$ {\mathcal V}_{\mathcal M}$ invariant, 
\begin{eqnarray}
\tilde{\mathcal{H}}\,{\mathcal V}_{\mathcal M}&\subseteq&
{\mathcal V}_{\mathcal M},
\label{subspc3}\\[4pt]
    {\mathcal V}_{\mathcal M}
&\eqdef&
\left\{\begin{array}{ll}
\mbox{Span}\left[1,x^2,\ldots,x^{2k},\ldots,x^{\mathcal 
    M}\right],& \mathcal{M}:\mbox{even},\\[4pt]
\mbox{Span}\left[x,x^3,\ldots,x^{2k+1},\ldots,x^{\mathcal 
    M}\right],& \mathcal{M}:\mbox{odd},
\end{array}\right.
    \label{eq:V0def3}\\
&&\mbox{dim}{\mathcal V}_{\mathcal M}=\left\{
\begin{array}{ll}
{\mathcal M}/2+1,& {\mathcal M}: \mbox{even},\\
({\mathcal M}+1)/2,& {\mathcal M}: \mbox{odd},
\end{array}
\right.
\label{Vdimform3}
\end{eqnarray}
but $\tilde{\mathcal{H}}$ is no longer a  {\em tri-diagonal\/} 
 matrix, 
$(\tilde{\mathcal{H}})_{j\,k}\neq0$, $j\ge k-1$.
The Hamiltonian ${\mathcal H}$ (\ref{factform}) is obviously hermitian
(self-adjoint) and all the eigenvalues are real 
and eigenfunctions can be chosen
real.
 We can obtain a finite number of exact eigenvalues and 
eigenfunctions by sweeping in a similar way as in the sextic potential case
(\ref{eigfun}), (\ref{Vdimform}). The oscillation theorem linking the number
of eigenvalues (from the ground state) to the zeros of eigenfunctions
does not hold in the difference equations.
The square integrability of all
the eigenfunctions $\int_{-\infty}^\infty\phi^2(x)dx<\infty$ is obvious.

\subsubsection{Difference Equation Analogue of Harmonic Oscillator Deformed by Sextic
Potential II}
Another difference analogue of the sextic potential Hamiltonian (\ref{sextic}) has
the same form as (\ref{H1}), (\ref{factform}) and (\ref{defAAdag}),
with only the potential function $V(x)$ and the compensation term
$\alpha_\mathcal{M}(x)$ are different:
\begin{eqnarray}
 V(x)&\eqdef&(a+i x)(b+i x)V_0(x),\quad 
V_0(x)\eqdef(c+ ix)(d+i x),\quad 
a,b,c,d\in\mathbb{R}_+,\\
\alpha_\mathcal{M}(x)&\eqdef& \mathcal{M}\left(\mathcal{M}-1+2(a+b+c+d)\right)x^2.
\end{eqnarray}
This Hamiltonian can be considered as a deformation by a  quadratic polynomial
factor $(a+i x)(b+i x)$ of the exactly solvable `discrete' quantum mechanics
having the continuous Hahn polynomials as  eigenfunctions \cite{os4os5}, another
difference analogue of the harmonic oscillator. 
See the comments in section 5 of \cite{os10}.

The  {\em pseudo ground state\/} wavefunction $\phi_0(x)$ annihilated by the $A$
operator $A\phi_0=0$ reads
\begin{eqnarray}
&&\phi_0(x)\\
&&\eqdef\sqrt{\Gamma(a+i x)\Gamma(a-i x)
\Gamma(b+i x)\Gamma(b-i x)\Gamma(c+ix)\Gamma(c-i x)\Gamma(d+ix)\Gamma(d-i x)}.
\nonumber
\end{eqnarray}
Again it has no node and it is square integrable.
The similarity transformed Hamiltonian acting on the polynomial space is
\begin{eqnarray}
 \tilde{\mathcal{H}}&\eqdef&
  \phi_0^{-1}\circ\mathcal{H}\circ\phi_0
  =V(x)\left(e^{-i\partial_x}-1\right)+V(x)^*\left(e^{i\partial_x}-1\right)
\nonumber\\
  &&\qquad\qquad\qquad\qquad+\mathcal{M}\left(\mathcal{M}-1+2(a+b+c+d)\right)x^2.
  \label{tilH3}
\end{eqnarray}
It is straightforward to verify the relationship (\ref{Hactx}) and 
to establish the
existence of the invariant polynomial subspaces of given parity (\ref{subspc3}), 
(\ref{eq:V0def3}) and (\ref{Vdimform3}).
The hermiticity of the Hamiltonian and the square integrability of the
eigenfunctions also hold. Thus another example of quasi exactly solvable 
difference equation is established.
\section{Other Examples}
\label{Examples}
\setcounter{equation}{0}

The other two examples are the difference equation analogues of the harmonic
oscillator with the centrifugal potential deformed by the sextic potential.
There are two types corresponding to the linear and quadratic polynomial
deformations. The corresponding exactly solvable difference equation has the Wilson
polynomial \cite{os4os5,And-Ask-Roy,koeswart} as the eigenfunctions.
 The last example is the difference analogue of the model discussed in
subsection \ref{cos2mod}, $1/\sin^2x$ potential deformed by
a $\cos2x$ potential. In this case the corresponding exactly solvable difference equation 
has the Askey-Wilson
polynomials \cite{os4os5,And-Ask-Roy,koeswart} as  eigenfunctions.
The basic idea for showing quasi exact solvability is almost the same as shown
above. 

\subsection{Difference Equation Analogues of Harmonic Oscillator With
Centrifugal Potential Deformed by Sextic Potential}
\label{difsex}

The Hamiltonians  have
the same form as (\ref{H1}), (\ref{factform}) and (\ref{defAAdag}),
with only the potential function $V(x)$ and the compensation term
$\alpha_\mathcal{M}(x)$ are different:
\begin{eqnarray}
\mbox{Type I}:\ V(x)&\eqdef&(b+i x)V_0(x),\qquad
\alpha_\mathcal{M}(x)\eqdef 
\mathcal{M}x^2,\\
\mbox{Type II}:\ V(x)&\eqdef&(a+i x)(b+i x)V_0(x),\\
\alpha_\mathcal{M}(x)&\eqdef&
\mathcal{M}\left(\mathcal{M}-1+(a+b+c+d+e+f)\right)x^2,
\end{eqnarray}
with a common $V_0(x)$
\begin{eqnarray}
V_0(x)&\eqdef&\frac{(c+ ix)(d+i x)(e+i x)(f+i x)}{2ix(2ix +1)},
\quad a,b,c,d,e,f\in\mathbb{R}_+-\{{1}/{2}\}.
\label{V0Wilson}
\end{eqnarray}
None of the parameters $a$, $b$, $c$, $d$, $e$ or $f$ should take the value
$1/2$, since it would cancel the denominator.
Because of the centrifugal barrier, the dynamics is constrained to a half line;
$0<x<\infty$. The type I case can also be considered as a quadratic polynomial
deformation of  the exactly solvable dynamics with  $V_{01}(x)$:
\begin{eqnarray}
\mbox{Type I}:\ V(x)&\eqdef&(b+i x)(c+i x)V_{01}(x),\quad
V_{01}(x)\eqdef\frac{(d+i x)(e+i x)(f+i x)}{2ix(2ix +1)},
\label{V0condualHahn}
\end{eqnarray}
which has the continuous dual Hahn polynomials \cite{os4os5,And-Ask-Roy,koeswart} as
eigenfunctions. This re-interpretation does not change the dynamics, since the
Hamiltonian and $A$ and $A^\dagger$ operators depend on $V(x)$.

The {\em pseudo ground state\/} wavefunction $\phi_0(x)$ is determined as
the zero mode of  the
$A$ operator $A\phi_0=0$:
\begin{eqnarray}
\mbox{Type I}:\ \phi_0(x)&\eqdef& \frac{
\sqrt{\prod_{j=2}^6 \Gamma(a_j+i x)\Gamma(a_j-i x)}}{\sqrt{\Gamma(2i
x)\Gamma(-2ix)}},\\
\mbox{Type II}:\ \phi_0(x)&\eqdef& \frac{
\sqrt{\prod_{j=1}^6\Gamma(a_j+i x)\Gamma(a_j-i x)}}{\sqrt{\Gamma(2i
x)\Gamma(-2ix)}},
\end{eqnarray}
in which the numbering of the parameters
\begin{equation}
a_1\eqdef a,\ a_2\eqdef b,\ a_3\eqdef c,\ a_4\eqdef d,\ a_5\eqdef e,\ 
a_6\eqdef f, 
\end{equation}
is used. It is obvious that both $\phi_0$ have no node in the half line
$0<x<\infty$. 

The similarity transformed Hamiltonian acting on the polynomial space has the same
form as before (\ref{tilH3})
\begin{eqnarray}
 \tilde{\mathcal{H}}&\eqdef&
  \phi_0^{-1}\circ\mathcal{H}\circ\phi_0
  =V(x)\left(e^{-i\partial_x}-1\right)+V(x)^*\left(e^{i\partial_x}-1\right)
+\alpha_\mathcal{M}(x).
  \label{tilH4}
\end{eqnarray}
Although the potential $V(x)$ has the harmful looking denominator
$1/\{2ix(2ix+1)\}$,
it is straightforward to verify that $\tilde{\mathcal{H}}$ maps a polynomial in
$x^2$ into another:
\begin{equation}
\tilde{\mathcal{H}}\,x^{2n}=\left\{\begin{array}{lll}
\sum_{j=0}^{n+1}a_{n,\,j}x^{2n+2-2j},& n\leq \mathcal{M}-1,&
a_{n,\,j}\in\mathbb{R},\\[6pt]
\sum_{j=0}^{\mathcal{M}}a'_{n,\,j}x^{2\mathcal{M}-2j},&  n=
\mathcal{M},&a'_{n,\,j}\in\mathbb{R}.
\end{array}\right.
\label{Hactx2}
\end{equation}
This is because $V_0$, which has the above denominator, keeps the polynomial
subspace of any even degree invariant, reflecting the exact solvability. 
This establishes that $\tilde{\mathcal{H}}$ keeps the polynomial space
$ {\mathcal V}_{\mathcal M}$ invariant, 
\begin{eqnarray}
\tilde{\mathcal{H}}\,{\mathcal V}_{\mathcal M}&\subseteq&
{\mathcal V}_{\mathcal M},
\label{subspc4}\\[4pt]
    {\mathcal V}_{\mathcal M}
&\eqdef&
\mbox{Span}\left[1,x^2,\ldots,x^{2k},\ldots,x^{2\mathcal 
    M}\right],\quad
\mbox{dim}{\mathcal V}_{\mathcal M}=
{\mathcal M}+1.
\label{subspc5}
\end{eqnarray}
The hermiticity of the Hamiltonians is obvious and the square-integrability
of the eigenfunctions $\int_{0}^\infty\phi^2(x)dx<\infty$ holds true.
This establishes the quasi exact solvability.

The corresponding quantum mechanical system has the prepotential $W(x)$ 
and the compensation term $\alpha_\mathcal{M}(x)$ as
\begin{eqnarray}
W(x)&=&-\frac{a}{4}x^4-\frac{b}{2}x^2+g\log x,\qquad a,\ b, \ g>0,
\quad 0< x<\infty,
\label{laguerW}\\ 
\alpha_\mathcal{M}(x)&=&-a\mathcal{M}x^2.
\end{eqnarray}
This and the above two difference analogue systems share the same invariant
polynomial subspace (\ref{subspc4}), (\ref{subspc5}). 
The undeformed exactly solvable system, 
{\em i.e.\/} (\ref{laguerW}) with $a=0$,
has the Laguerre polynomials as  eigenfunctions. The corresponding 
undeformed exactly solvable difference equations determined by 
$V_0$ (\ref{V0Wilson}), $V_{01}$ (\ref{V0condualHahn}) have the Wilson and
the continuous dual Hahn polynomials as  eigenfunctions.
These are three and two parameter deformation of the Laguerre polynomial
\cite{os4os5,And-Ask-Roy,koeswart}.
\subsection{Difference Equation Analogue of $1/\sin^2x$ Potential Deformed by
$\cos2x$ Potential}
\label{diftrig}

This system is a quasi exactly solvable deformation of the exactly solvable
dynamics which has the Askey-Wilson polynomials \cite{os4os5,And-Ask-Roy,koeswart}
as  eigenfunctions. Let us first introduce the variables and notation
appropriate for the Askey-Wilson polynomials. 
We use variables $\theta$, $x$ and $z$,
which are related as
\begin{equation}
  0<\theta<\pi,\quad x=\cos\theta,\quad z=e^{i\theta}.
\end{equation}
The dynamical variable is $\theta$ and the inner product 
 is
$\langle f|g\rangle=\int_0^{\pi}d\theta f(\theta)^*g(\theta)$.
We denote $D\eqdef z\frac{d}{dz}$. Then $q^D$ is a $q$-shift
operator, $q^Df(z)=f(qz)$, with $0<q<1$. 
The Hamiltonian is obtained by deforming the potential function $V_0(z)$ by a
linear polynomial in $z$:
\begin{eqnarray}
  \mathcal{H}&\eqdef&\sqrt{V(z)}\,q^{D}\!\sqrt{V(z)^*}
  +\sqrt{V(z)^*}\,q^{-D}\!\sqrt{V(z)}
  -(V(z)+V(z)^*)+\alpha_\mathcal{M}(z),
  \label{H-q}\\[2pt]
&=&A^\dagger A+\alpha_\mathcal{M}(z),\qquad\qquad
\alpha_\mathcal{M}(z)\eqdef -abcde q^{-1}(1-q^\mathcal{M})(z+\frac{1}{z}),
\\[2pt]
{A}^{\dagger}&\eqdef&
  -i\left(\sqrt{V(z)}\,q^{\frac{D}{2}}
  -\sqrt{V(z)^*}\,q^{-\frac{D}{2}}\right),\
{A}\eqdef
  i\left(q^{\frac{D}{2}}\sqrt{V(z)^*}
  -q^{-\frac{D}{2}}\sqrt{V(z)}\right),\\[2pt]
V(z)&\eqdef&(1-a z)V_0(z),\quad 
V_0(z)\eqdef\frac{(1-bz)(1-cz)(1-dz)(1-ez)}{(1-z^2)(1-qz^2)},\\
&&\qquad\qquad\qquad\qquad -1<a,b,c,d,e<1.
\end{eqnarray}

The {\em pseudo ground state\/} wavefunction $\phi_0(z)$ is determined as
the zero mode of  the
$A$ operator $A\phi_0=0$:
\begin{equation}
 \phi_0(z)\eqdef
\sqrt{\frac{(z^2,z^{-2};q)_{\infty}}
  {(az,az^{-1},bz,bz^{-1},cz,cz^{-1},dz,dz^{-1},ez,ez^{-1};q)_{\infty}}},
\end{equation}
where $(a_1,\cdots,a_m;q)_{\infty}=\prod_{j=1}^m
\prod_{n=0}^{\infty}(1-a_jq^n)$.
Obviously $\phi_0$ has no node or singularity in $0<\theta<\pi$.
We look for exact eigenvalues and eigenfunctions
of the  Hamiltonian (\ref{H-q}) in the form:
\begin{equation}
{\mathcal H}\phi={\mathcal E}\phi,\quad
\phi(z)=\phi_0(z)P_{\mathcal M}(x),
\label{eigfunz}
\end{equation}
in which $P_{\mathcal M}(x)$ is a degree $\mathcal{M}$ polynomial in $x$ or in
$z+{1}/{z}=2\cos\theta=2x$.
The similarity transformed Hamiltonian acting on the polynomial space 
has the form 
\begin{equation}
  \tilde{\mathcal{H}}\eqdef
  \phi_0^{-1}\circ \mathcal{H}\circ\phi_0
  =V(z)\left(q^{D}-1\right)+V(z)^*\left(q^{-D}-1\right)
  -abcde q^{-1}(1-q^\mathcal{M})(z+\frac{1}{z}).
  \label{tildeHAS}
\end{equation}
Without the deformation factor $1-az$ and the compensation term, the above
Hamiltonian $\tilde{\mathcal{H}}$ keeps the polynomial subspace in $z+{1}/{z}$
of any degree invariant.
It is straightforward to show 
\begin{eqnarray}
\tilde{\mathcal{H}}\,{\mathcal V}_{\mathcal M}&\subseteq&
{\mathcal V}_{\mathcal M},
\label{subspc6}\\[4pt]
    {\mathcal V}_{\mathcal M}
&\eqdef&
\mbox{Span}\left[1,z+\frac{1}{z},\ldots,\left(z+\frac{1}{z}\right)^k,\ldots,
\left(z+\frac{1}{z}\right)^\mathcal{M}\right],\
\mbox{dim}{\mathcal V}_{\mathcal M}=
{\mathcal M}+1.
\label{subspc7}
\end{eqnarray}
The hermiticity of the Hamiltonian is obvious and the square-integrability
of the eigenfunctions  holds also true.
This establishes the quasi exact solvability.
\section{Summary  and Comments}
\setcounter{equation}{0}

First let us summarise:
Five explicit examples of quasi exactly solvable difference equations of one
degree of freedom are derived by multiplicatively deforming the
the known exactly solvable difference equations for the Meixner-Pollaczek,
continuous Hahn, continuous dual Hahn, Wilson and Askey-Wilson polynomials.
The finite dimensional Hamiltonian matrix, no longer tri-diagonal, can be solved
exactly by sweeping.
All the eigenvalues and eigenfunctions are real, but the oscillation theorem,
connecting the excitation level to the number of zeros, does not hold.
Similarity and contrast with the known QES examples in ordinary quantum mechanics;
the harmonic oscillator (with or without the centrifugal potential) deformed by a
sextic potential and the $1/\sin^2x$ potential deformed by a $\cos2x$ potential,
are demonstrated in some detail.

A few comments are in order.
 First let us stress
that the mere existence of the finite dimensional invariant  polynomial subspace
$\tilde{\mathcal{H}}\,{\mathcal V}_{\mathcal M}\subseteq {\mathcal V}_{\mathcal
M}$ is not sufficient for the quasi exact solvability. The theory must be endowed
with a {\em pseudo ground state\/} wavefunction
$\phi_0$, which must be {\em nodeless\/} and square integrable.
Moreover, the reverse similarity transformed Hamiltonian 
$\mathcal{H}\eqdef
  \phi_0\circ \tilde{\mathcal{H}}\circ\phi_0^{-1}$ must be hermitian,
in order to guarantee the real spectrum.

Let us consider an {\em additively deformed\/} potential with a cubic
term
\begin{equation}
V(x)\eqdef a x^3+V_0(x),\quad 
\alpha_\mathcal{M}(x)\eqdef a\mathcal{M}(\mathcal{M}-1)x,\quad
a\in\mathbb{R},
\end{equation}
with an exactly solvable $V_0$, for example,
\begin{equation}
V_0(x)=b+ix,\quad \mbox{or}\quad (b+ix)(c+ix),\quad b, c\in \mathbb{R}_+,
\end{equation}
corresponding to the Meixner-Pollaczek and the continuous Hahn polynomials.
It is rather trivial to verify the existence of 
the invariant polynomial subspace:
\begin{eqnarray}
\tilde{\mathcal{H}}&\eqdef&
  V(x)\left(e^{-i\partial_x}-1\right)+V(x)^*\left(e^{i\partial_x}-1\right)
+a\mathcal{M}(\mathcal{M}-1)x,
\label{subspc8}\\
\tilde{\mathcal{H}}\,{\mathcal V}_{\mathcal M}&\subseteq&
{\mathcal V}_{\mathcal M},\quad
{\mathcal V}_{\mathcal M}\eqdef
\mbox{Span}\left[1,x,x^2,\ldots,x^{k},\ldots,x^{\mathcal 
    M}\right].
\label{subspc9}
\end{eqnarray}
Although we have not been able to derive the explicit form of the
pseudo ground state wavefunction $\phi_0(x)$ as a solution of $A\phi_0=0$,
it seems rather unlikely that the $\phi_0$ satisfies the above mentioned criteria.
This is because the corresponding quantum mechanical case,
the quartic oscillator,
\begin{equation}
W(x)=ax^3-\frac{b}{2}x^2,\quad \alpha_{\mathcal M}(x)=6a{\mathcal M}x,
\end{equation}
also have the invariant polynomial subspace (\ref{subspc9}).
The square integrability of $\phi_0(x)=e^{W(x)}$ does not hold whichever sign $a$
might take.

Let us discuss the hermiticity of the Hamiltonians (\ref{H1}) and (\ref{H-q}).
The hermiticity means $\langle g|\mathcal{H}f\rangle=\langle \mathcal{H}g|f\rangle$
for a given inner product $\langle g|f\rangle$ for arbitrary elements $f$ and $g$
in a certain dense subspace of the appropriate Hilbert space.
The obvious choice for such a subspace is spanned by the `pseudo 
ground state' wavefunction $\phi_0$
times polynomials. The types of the polynomials are:
\begin{eqnarray}
&&(a): \mbox{polynomials in } x\ \mbox{for the Hamiltonians in section}\ \ref{dqmex},\\
&&\quad  \langle g|f\rangle=\int_{-\infty}^\infty g(x)^*f(x)dx,\quad f(x)=\phi_0(x)P(x),
\quad g(x)=\phi_0(x)Q(x),\\
&&(b):\mbox{polynomials in } x^2\ \mbox{for the Hamiltonians in section}\ \ref{difsex},\\
&&\quad  \langle g|f\rangle=\int_{0}^\infty g(x)^*f(x)dx,\quad f(x)=\phi_0(x)P(x^2),
\quad g(x)=\phi_0(x)Q(x^2),\\
&&(c): \mbox{polynomials in } x=\cos\theta\ \mbox{for the Hamiltonians 
in section}\ \ref{diftrig},\\
&&\quad  \langle g|f\rangle=\int_{0}^\pi g(\theta)^*f(\theta)d\theta,
\quad f(\theta)=\phi_0(z)P(\cos\theta),
\quad g(\theta)=\phi_0(z)Q(\cos\theta).
\end{eqnarray}

The Hamiltonians (\ref{H1}) and (\ref{H-q}) consist of three parts:
\begin{equation}
\mathcal{H}=\mathcal{H}_1+\mathcal{H}_2+\mathcal{H}_3,\quad
\mathcal{H}_3= \alpha_\mathcal{M}-(V+V^*),
\end{equation}
and
\begin{eqnarray}
\mathcal{H}_1&=&\sqrt{V(x)}\,e^{-i\partial_x}\sqrt{V(x)^*}, \quad
\mathcal{H}_2=\sqrt{V(x)^*}\,e^{i\partial_x}\sqrt{V(x)}, \quad
\mbox{for}\  (\ref{H1}),\\
\mathcal{H}_1&=&\sqrt{V(z)}\,q^{D}\!\sqrt{V(z)^*}, \quad\ \ \
\mathcal{H}_2=\sqrt{V(z)^*}\,q^{-D}\!\sqrt{V(z)}, \quad\  
\mbox{for}\  (\ref{H-q}).
\end{eqnarray}
It is obvious that $\mathcal{H}_3$ is hermitian by itself. 
When $\mathcal{H}_1$ acts on $f$, the argument is shifted from $x$ to $x-i$ or from $\theta$ to
$\theta-i\log q$. With the compensating change of integration variable 
from $x$ to $x+i$ or from $\theta$ to
$\theta+i\log q$ one can formally show 
$\langle g|\mathcal{H}_1f\rangle=\langle \mathcal{H}_1g|f\rangle$ in a straightforward way.
Similarly we have $\langle g|\mathcal{H}_2f\rangle=\langle \mathcal{H}_2g|f\rangle$
by another change of integration variable.
This is the `formal hermiticity.'

Actually, the shift of integration variable, to be realised by the Cauchy integral,
would involve additional integration contours:
\begin{eqnarray}
&&(a): (-\infty,\pm i-\infty),\quad (+\infty,\pm i+\infty)\quad 
\mbox{for the Hamiltonians in section}\ \ref{dqmex},\label{1type}\\
&&(b): (0,\pm i),\quad (+\infty,\pm i+\infty)\qquad\quad  \mbox{for the Hamiltonians in section}\ \ref{difsex},\\
&&(c): (0,\pm i\log q),\quad (\pi,\pi \pm i\log q)\quad \mbox{for the Hamiltonians 
in section}\ \ref{diftrig}.
\label{3type}
\end{eqnarray}
It should be noted that all the singularities arising from $V$ and $V^*$ in cases 
($b$) and ($c$) are cancelled by the zeros coming from the pseudo 
ground state wavefunctions $\phi_0$ and $\phi_0^*$, and the Cauchy integration formula applies 
in all cases. The contribution of the additional contour integrals (\ref{1type})--(\ref{3type})
cancel with each other and the shifts of integration variables is justified.

To be more specific, the contribution from the contours at  infinity in ($a$) and ($b$)  
vanish identically due to the strong damping by $\phi_0$ and $\phi_0^*$.
The contribution from the two vertical contours in  ($b$) passing the origin 
cancel with each other due to the evenness of $\phi_0$, $\phi_0(x)=\phi_0(-x)$ and the
polynomials in $x^2$, $P((-x)^2)=P(x^2)$. 
Likewise the contribution from the two vertical contours in  ($c$) passing the origin
cancel with each other.
Those in  ($c$) passing $\pi$ also
cancel with each other due to the $2\pi$ periodicity of $\phi_0$, 
$\phi_0(\theta)=\phi_0(2\pi+\theta)$.
Note that the hermiticity in ($b$) and ($c$)  cases holds only for the sum
$\mathcal{H}_1+\mathcal{H}_2$. This concludes the comments on hermiticity.

It is now evident that the scope of the present method,
deformation of $W$ or $V$ for generating a quasi exactly solvable system
from an exactly solvable dynamics, is rather limited.
It is highly unlikely to get a quasi exactly solvable system, if $W$ contains a term
higher than $x^5$ or $\cos4x$, or if $V$ has a form $\prod_{j=1}^n(a_j+ix)$,
$n\ge5$ or $\left(\prod_{j=1}^n(a_j+ix)\right)/\{2ix(2ix+1)\}$, $n\ge7$.
However, generalisation to multi-particle difference equations,
the Ruijsenaars-Schneider-van Diejen systems \cite{RS,vD}, is possible \cite{os10}.
The corresponding ordinary quantum mechanical systems, {\em i.e.\/}
quasi-exactly solvable Calogero-Sutherland systems \cite{cal,sut} are derived by
Sasaki and Takasaki \cite{st1}.

For simplicity of presentation, we have restricted the parameters $a$, $b$,
$c$, $d$, $e$ and $f$ in $V(x)$ to be real, positive etc.
In most cases the ranges of parameters can be relaxed without losing quasi exact
solvability. For example, in the first example of subsection \ref{difex1}, a
configuration
$a>0$ and $b$ and $c$ are complex conjugate, $b=c^*$ with positive real parts,  
is also possible.

It is a challenge to see if these newly derived quasi exactly solvable difference
equations  can be understood by the existing
ideas of QES; the $sl(2)$ algebra \cite{turb} and/or  its deformations, various
generalised super symmetry ideas \cite{N-SUSY}.
Can we use the Bethe ansatz method to solve these problems?
Are these QES systems equivalent to some spin systems
\cite{ulyanov}?

\section*{Acknowledgements}

We thank Choon-Lin Ho for useful comments.
This work is supported in part by Grants-in-Aid for Scientific
Research from the Ministry of Education, Culture, Sports, Science and
Technology, No.18340061 and No.19540179.


\end{document}